# Phononic thermal resistance due to a finite periodic array of nano-scatterers

T.T. Trang Nghiêm and P.-Olivier Chapuis

*CETHIL-UMR5008, INSA de Lyon and CNRS, UMR 5008, 69621 Villeurbanne, France*

The wave property of phonons is employed to explore the thermal transport across a finite periodic array of nano-scatterers such as circular and triangular holes. As thermal phonons are generated in all directions, we study their transmission through a single array for both normal and oblique incidences, using a linear dispersionless time-dependent acoustic frame in a two-dimensional system. Roughness effects can be directly considered within the computations without relying on approximate analytical formulae. Analysis by spatio-temporal Fourier transform allows us to observe the diffraction effects and the conversion of polarization. Frequency-dependent energy transmission coefficients are computed for symmetric and asymmetric objects. We demonstrate that the phononic array acts as an efficient thermal barrier by applying the theory of thermal boundary (Kapitza) resistances to arrays of smooth scattering holes in silicon for an exemplifying periodicity of 10 nm in the [5-100 K] temperature range. It is observed that the associated thermal conductance has the same temperature dependence than that without phononic filtering.

## I. INTRODUCTION

Characterizing heat flow at interfaces [1, 2], where thermal boundary resistances (TBR, also called Kapitza resistances) can impede the heat transfer, is of critical importance at the sub-micron length scales due to the high surface-to-volume ratio. Adding carbon nanotubes to the metal/metal interface [3, 4], or including self-assembled nanoparticles embedded in the material [5] [6] are some experimental examples that can be used to tune the TBR, as well as chemical etching [7-9]. Other strategies have been highlighted, such as the addition of a new material to the solid/solid coupling or modulating the roughness [10]. Lattice dynamics [11, 12], Green's functions [13] and molecular dynamics are often employed for the calculations [14-16]. We observe that the two commonly applied models, the acoustic mismatch model (AMM) and diffuse mismatch model (DMM), do not necessarily lead to values that are comparable to the available experimental data [1, 15, 17, 18]. In all cases, it is clear that the shapes of embedded elements and/or interfaces have an important impact on the TBR [5], [6], [10] [19], [20].

Phonon coherence effects may provide a new way to control heat transfer properties at boundaries. Such effects linked to the wave nature of the phonons have recently been experimentally evidenced by the Chen group [21] and the California group [22], showing in particular that peculiar effects may take place for nanometer sizes at low temperatures and up to temperatures close to ambient. In addition, phononic configurations, which involve periodic arrays of holes [23] [24] [25] or embedded particles [26], have shown a strong reduction of the effective thermal conductivity. It is debated if the reason is due to coherence in these particular experimental works or due to the involved sizes which appear larger than the thermal wavelengths, but many interesting proposals have been highlighted, from the GHz to the THz [27, 28] range where periodicity is expected to play a key role for heat transport [23, 29]. Most of the theoretical suggestions deal with crystals involving many periods and derive an effective thermal conductivity based on the infinite-crystal approach. Since boundaries can already decrease strongly the heat transfer, it may be possible to block the heat transfer with smaller structures.

Here, we study the transmission and the thermal resistance of a finite phononic structure consisting of a single array of periodic holes, by solving the elastic wave equation. The acoustic frame is particularly suitable to reproduce the low-temperature phonon behaviors [30] and may help to disentangle the elastic and inelastic contributions to thermal resistances [31]. Various hole shapes are considered, such as disks, equilateral and isosceles triangles. We note that roughness can be directly included in the model by designing associated shapes. In contrast to previous works that analyzed phonon transmission through such single array with the goal of highlighting non-symmetrical acoustic transmission (sometimes improperly called 'acoustic rectification' [32]), we analyze the acoustic transmission not only for the normal incidence but also for oblique cases. This is required because thermal phonons are generated randomly in all directions. It allows observing reciprocity also for the case of asymmetric phononic structures. We perform an analysis of the displacement fields in order to determine if phonons of particular wavelengths and direction of propagation are especially filtered by the single array. Finally, we calculate the thermal conductances of the phononic array with the help of a Landauer-based approach which is similar to the AMM theory of thermal boundary resistance, both at equilibrium and out of equilibrium. A 10 nm-periodicity is considered for this example in two dimensions, showing that the structure indeed blocks a large portion of heat depending on the geometrical parameters associated to the considered shapes.

The article is organized as follows: In Sec. II, we describe (i) the structure, (ii) how we simulate phonon propagation with the elastic wave equation, (iii) the implementation and the computational domain, and (iv) present briefly the derivation of the thermal conductance based on the frequency-dependent transmission coefficients. In Sec. III, we (i) analyze the results of the spatial filtering resulting in diffraction effects and (ii) compute the transmission coefficients and the thermal conductance in the presence of the hole array. Finally, we present a summary and the consequences of this work in Sec. IV.

## II. Method
### A. Simulated structure and computational domain

The illustration of one simulated structure in two dimensions (2D) is shown in Figure 1. The periodicity in the $y$-direction is $a$ and we simulate often more than one period, as will be explained in Sec. II.C. Empty holes with specific shapes are located in the middle of the simulated domain. Periodic boundary conditions are applied at the bottom and the top of structure. Two absorbing layers are created to avoid the reflection at left and right of the system [33-35]. The size of these layers is large enough to ensure that the fluxes at the end walls of the system are null. We recall the approach that we use to derive the elastic equation in the absorbing domain in Appendix A.

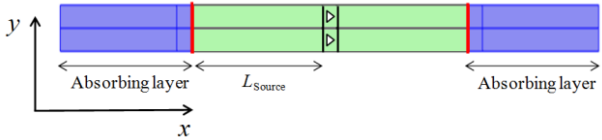

FIG 1. Illustration of a simulated structure involving two rows with triangular shapes. Absorbing conditions are applied in the two blue regions (see more in Appendix); phonon source and detector are highlighted with the red lines.

Acoustic waves are generated by the source at the left end of the lossless computational domain (left red line) and detected at its right end (right red line), and vice versa. The comparison between the propagation from left to right and right to left is particularly useful for the analysis of asymmetric objects. The distances between the single array and the source/detector are equal to $L_{source} = 10\lambda$, where $\lambda$ is the wavelength. The mesh is chosen so that both $\Delta x$ and $\Delta y$ are always smaller than $\lambda/5$. The wavelengths can be compared to the periodicity by introducing a non-dimensioned positive number $N_\lambda = a/\lambda$. Hence, the circular frequency is also related to the medium velocity $v$ and the periodicity $a$:

$$\omega = \frac{2\pi v}{a} N_\lambda \, , \qquad (1)$$

where $v$ is the velocity for the considered polarization. Three shapes of holes are analyzed in this study: (a) circular, (b) equilateral triangle, (c) right isosceles triangle. To compare the area of these holes, we introduce a "*filling factor*" $f$ defined by the ratio between the hole area and the $a$-side square area:

$$f = \frac{S_{hole}}{S_{square}} \, . \qquad (2)$$

This can be seen as the filling factor of the corresponding phononic crystal (of 2D periodicity). In addition, we consider what we will call the "*blocking ratio*" defined as the ratio of the projected length $l$ to the periodicity $a$:

$$r = \frac{l}{a} \, . \qquad (3)$$

Fig. 2 presents three hole shapes with the same filling factor $f = 0.2$, with blocking ratios which are different. The values of the blocking ratios increase with the following order: disk, equilateral triangle, and then isosceles triangle.

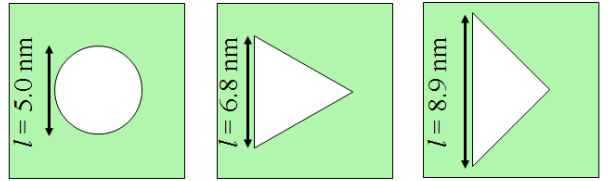

FIG 2. Three different hole shapes studied in this work: disk, equilateral triangle and right isosceles triangle, respectively from left to right. The periodicity is $a$ and the filling factor $f = 0.2$ is identical for these three cases.

### B. Phonons as elastic waves

As phonons are propagative acoustic waves, we solve the linear elastic equation in two dimensions (2D) to compute the phonon transmission experiment through an array:

$$\rho(\vec{r}) \frac{\partial^2 \vec{u}(\vec{r},t)}{\partial t^2} = \nabla T(\vec{r},t), \qquad (4)$$

where $\rho$ is the mass density, $\vec{u}$ is displacement field and $T$ is the stress tensor. The stress tensor relates to the elastic constant $C_{ijkl}$ by $T_{ij}(\vec{r},t) = \frac{1}{2} C_{ijkl}(\vec{r}) \left( \frac{\partial u_k(\vec{r},t)}{\partial x_l} + \frac{\partial u_l(\vec{r},t)}{\partial x_k} \right)$. In the following, we use the usual abbreviated subscripts for elastic constants [36] and require only three constants $C_{11}$, $C_{44}$ and $C_{12}$ for the cubic crystal case. To simplify, the material is assumed isotropic, so that $C_{12} = C_{11} - 2C_{44}$. For silicon, $\rho = 2331$ kgm$^{-3}$, $C_{11} = 16.57\times10^{10}$ Nm$^{-2}$ and $C_{44} = 7.956\times10^{10}$ Nm$^{-2}$ [36]. The longitudinal velocity $v_l$ and the transverse one $v_t$ are defined by the materials properties: $v_l = \sqrt{C_{11}/\rho}$ and $v_t = \sqrt{C_{44}/\rho}$.

We solve numerically Eq. 1 with the finite element method [37]. We obtain the time evolution of the displacement field $u(\vec{r},t)$ and the stress tensor $T_{ij}(\vec{r},t)$ at each point in the system. This allows computing the acoustic Poynting vector $\vec{P}(\vec{r},t)$ that carries the energy flux:

$$\vec{P}(\vec{r},t) = -\frac{\vec{v}^*(\vec{r},t)\hat{T}(\vec{r},t)}{2}, \qquad (5)$$

where $\vec{v}$ is the velocity calculated by derivation of the displacement field $d\vec{u}/dt$. In the following, we consider a bi-dimensional system and the Poynting vector is reduced to a vector with two components. The projection of the flux on the $x$-direction allows monitoring the energy propagation across the array:

$$P_x = \sum_{i=1,t} -\frac{v_x^*(\vec{r},t)T_{xx}(\vec{r},t) + v_y^*(\vec{r},t)T_{yy}(\vec{r},t)}{2}. \qquad (6)$$

We note that this method can be used to study phonon transmission for many types of structures, such as boundaries between two dissimilar solids in direct contact (see Appendix B). It may provide a mesoscopic alternative to atomistic techniques such as molecular dynamics, especially at low temperatures where the latter does not behave well. Note that it could also be extended to nonlinear media.

### C. Condition linking oblique waves and the simulated domain

As mentioned in Sec. I, a thermal source emits phonons to its surrounding environment in all directions. Hence, to calculate the heat flux generated and transmitted through a structure, all directions have to be included. We excite not only the normal acoustic waves that are perpendicular to the periodic direction, but also oblique waves (see Fig. 3). As longitudinal and transverse wave propagations can be separated [36] (see more in Sec. III.A), we show here how to simulate the longitudinal waves.

The acoustic wave displacement is expressed as

$$\vec{u} = \vec{u}_0 e^{i(\omega t - \vec{k}\cdot\vec{r})} = \vec{u}_0 e^{i(\omega t - k_x x - k_y y)}, \qquad (7)$$

where $\vec{k}$ is the wave vector, $k_x$ and $k_y$ are the projection of $\vec{k}$ on the $x$ and $y$ axis, respectively. $\vec{u}_0$ is perpendicular or parallel to $\vec{k}$, depending on the polarization. From Eq. (7), the acoustic source located in $\vec{r}_0$ writes $\vec{u}(\vec{r}_0) = \vec{u}_0 e^{i(\omega t - \vec{k}\cdot\vec{r}_0)}$. The periodic condition in the $y$-direction requires that at one given position $x_0$, the displacement field at the top and the bottom are the same. Fig. 3b shows two typical points $y_1$ and $y_2$ at the boundaries and illustrates the condition on the incident angle $\theta = (\vec{u},\vec{x})$ and the number of simulated rows $N_{rows}$. The condition is written as

$$u\left(y_1 = \frac{aN_{rows}}{2}, x_0\right) = u\left(y_2 = -\frac{aN_{rows}}{2}, x_0\right) \qquad (8)$$

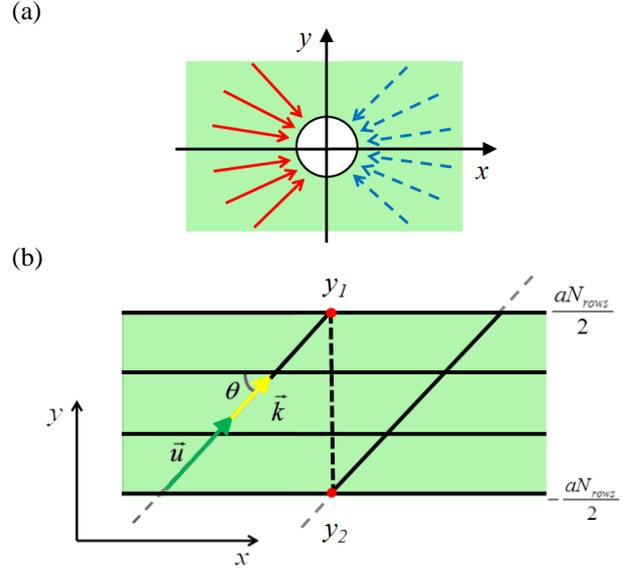

FIG 3. (a) Illustration of waves impinging the scatterer from different incident angles. (b) Periodic condition for the oblique incidence.

Taking Eq. 8 into account, we have

$$u_0 \exp\left[i\left(\omega t - k\sin\theta\frac{aN_{rows}}{2} - k\cos\theta\, x_0\right)\right]$$
$$= u_0 \exp\left[i\left(\omega t + k\sin\theta\frac{aN_{rows}}{2} - k\cos\theta\, x_0\right)\right] \qquad (9)$$

Eq. 9 leads easily to the relation

$$\sin\theta = \frac{2\pi v}{\omega a N_{rows}} n, \qquad (10)$$

where $n$ is an integer that allows to satisfy $0 \leq \sin\theta \leq 1$. By taking Eq. 1 into account, we obtain the final condition:

$$\sin\theta = \frac{n}{N_\lambda N_{rows}}. \quad (11)$$

As we consider angles $\theta$ from 0 to $\pi/2$, $n$ takes values between $0$ and $N_\lambda N_{rows}$. Consequently, for a given frequency $N_\lambda$ and a given number of simulated rows $N_{rows}$, we can simulate certain waves with incident angles satisfying the condition defined by Eq. 11.

Fig. 4 illustrates the propagation of the waves by presenting a snapshot of the $u_x$ component of the displacement field for two incident angles in the case $N_\lambda = 2.5$ (as an example, for $a = 10$ nm, $\lambda = a/2.5 = 4$ nm): (a) normal incident angle ($\theta = 0$), and (b) oblique incident angle with $\sin\theta = 0.4$. In this example the stationary regime has not been reached. The scattering objects separate the displacement field into two regions: a first one at left where incident and reflected waves are observed; a second one at right that contains the fields associated to the transmitted waves, where the propagation has been obviously modified. In addition, the interference between incident and reflected waves is observed in the first region. We note in particular that a $u_y$ component is generated immediately after the objects even though only the $u_x$ component is excited in the normal incident case (not shown here). This is due to the fact that the propagation direction is changed and is not anymore only in the normal incidence for longitudinal waves; the projection of the displacement field on the $y$-axis is non-null. Animations related to normal and oblique incidence of phonons on the single array during 313 wave periods can be found in [38].

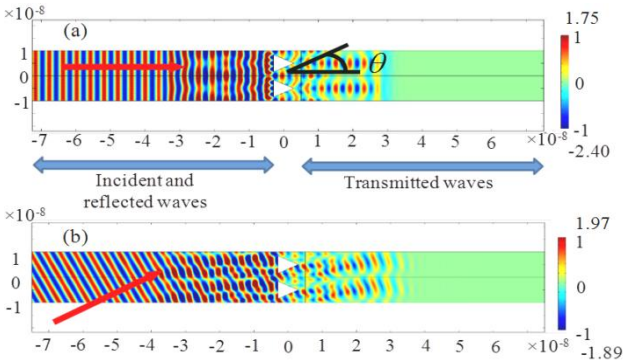

FIG 4. $u_x$ displacement field close to a single array of equilateral triangular holes: (a) wave in normal incidence, and (b) wave in normal incidence with $\sin\theta = 0.4$. Two regions are defined: the one with incident and reflected (IR) waves, and the one with transmitted (T) waves.

As waves travel through a grating-like periodic structure, they can be diffracted according to Bragg's law expressed as

$$a\sin\theta_n = n\lambda, \quad (12)$$

where $n$ is an integer which corresponds to the diffraction order characterized by the angle $\theta_n$ (see also Appendix C). Considering the wave shown in Fig. 4 with $N_\lambda=2.5$, the diffraction angles are $\theta_1 = 1/2.5 = 23.6°$ and $\theta_2 = 2/2.5 = 53.1°$. In the following, the magnitude of these waves will be determined by the spatio-temporal Fourier transform.

### D. Equilibrium thermal conductance in 2D and frequency-dependent transmission coefficients

The TBR (Kapitza) measures the boundary resistance to the propagation of the thermal flux. The theory has been applied to predict the thermal resistance of different types of junction [15, 16, 39, 40]. Here, we consider the hole array as a barrier between two parts of same material. In the following, we recall how to apply the formula for the 2D case. The thermal conductance $G$ between two media is defined as:

$$G = \frac{q}{T_2 - T_1} \quad (13)$$

where $q$ is the heat flux across the junction, and $T_i$ is the temperatures at the lead $i$. The heat flux across the structure relates to the transmitted phonons. At steady-state, the net heat flux in 2D is

$$q = \int_{\vec{k}}^{+} \hbar\omega v_{1x}(\omega,\vec{k}) t_{12}(\omega,\vec{k}) f(\vec{k},T_1) \frac{d^2\vec{k}}{(2\pi)^2}$$
$$+ \int_{\vec{k}}^{-} \hbar\omega v_{2x}(\omega,\vec{k}) t_{21}(\omega,\vec{k}) f(\vec{k},T_2) \frac{d^2\vec{k}}{(2\pi)^2} \quad (14)$$

where $\hbar$ is the reduced Plank constant, $v_{ix}$ denotes the phonon velocity in medium $i$ projected along the direction $x$ normal to the array, $f(\vec{k},T_i)$ is the phonon distribution function at the medium temperature $T_i$, $t_{ij}(\omega,\vec{k})$ is the wave-vector dependent transmission coefficient from the medium $i$ to the medium $j$. The signs $+$ and $-$ indicate that the integrals deal with $k_x > 0$ and $k_x < 0$, respectively. When the thermal transport is close to the equilibrium state, the phonon distribution can be assimilated to the equilibrium Bose-Einstein distribution that does not anymore depend on the direction:

$$f_{BE}(\omega,T) = \left[\exp\left(\frac{\hbar\omega}{k_B T}\right) - 1\right]^{-1}, \quad (15)$$

where $k_B$ is the Boltzmann constant. In a lossless medium, by invoking the balance principle when the heat flux is zero at thermal equilibrium [15], Eq. 14 can be simplified to

$$q = \int_{\vec{k}}^{+} \hbar \omega v_x(\omega, \vec{k}) t_{12}(\omega, \vec{k}) [f_{BE}(\omega, T_2) - f_{BE}(\omega, T_1)] \frac{d^2 \vec{k}}{(2\pi)^2}. \quad (16)$$

In 2D coordinates, $d^2\vec{k} = dk \cdot k \, d\theta = \frac{\omega}{v^2} d\omega \, d\theta$, where $\theta$ is the angle between $\vec{k}$ and the $x$ axis, $\theta = \left[-\frac{\pi}{2}, \frac{\pi}{2}\right]$ and therefore $v_x = v\cos\theta$. We can perform the integration on $\theta = \left[0, \frac{\pi}{2}\right]$ and the 2D thermal conductance at equilibrium $G = \frac{dq}{dT}$ associated to one polarization state is finally calculated as

$$\begin{aligned}G_{eq} &= 2 \frac{1}{(2\pi)^2} \int_\omega \int_{\theta=0}^{\theta=\frac{\pi}{2}} \frac{\hbar \omega^3}{v^2} \frac{d f_{BE}(\omega, T)}{dT} t_{12}(\omega, \theta) \cos\theta \, d\theta \, d\omega \\ &= \frac{1}{(2\pi)^2} \int_\omega \frac{\hbar \omega^3}{v^2} \frac{df_{BE}(\omega, T)}{dT} d\omega \times 2 \int_{\theta=0}^{\theta=\pi/2} t_{12}(\omega, \theta) \cos\theta \, d\theta.\end{aligned} \quad (17)$$

We introduce the frequency-dependent transmission coefficient $\tau_{12}(\omega)$ as the transmission coefficient including all wavevector directions:

$$\tau_{12}(\omega) = 2\int_{\theta=0}^{\theta=\pi/2} t_{12}(\omega, \theta) \cos\theta \, d\theta, \quad (18)$$

so that e 2D thermal conductance writes:

$$G_{eq} = \frac{1}{(2\pi)^2} \int_\omega \frac{\hbar \omega^3}{v^2} \frac{d f_{BE}(\omega, T)}{dT} \tau_{12}(\omega) \, d\omega. \quad (19)$$

The maximal monochromatic thermal conductance ($\tau_{12} = 1$) for one polarization state is then given by

$$g_{\omega, eq, \max} = \frac{1}{(2\pi)^2} \frac{\hbar \omega^3}{v^2} \frac{df_{BE}(\omega, T)}{dT}. \quad (20)$$

The expression of Eq. 17 is established at equilibrium. Out of equilibrium, a discontinuity of temperature occurs, which can be accounted for by using the local distribution $f(\vec{r})$ [39], which obeys the Boltzmann transport equation (BTE) with

$$f(\vec{r}) = f_{BE}(T(\vec{r})) + \delta f(\vec{r}). \quad (24)$$

Including the deviation of the distribution function $f$ in the relaxation time approximation of the BTE, the non-equilibrium thermal conductance can finally be written as [39]

$$G_{neq} = \frac{G_{eq}}{1 - \beta_{12} - \beta_{21}} \quad (25)$$

involving the quantity

$$\beta_{12} = \int_{\vec{k}}^{+} \tau_{scatt1} v_{1x}^2 \hbar \omega \frac{\partial f_{eq}}{\partial T} t_{12} \, d\vec{k} / \kappa_1 \quad (26)$$

where $\tau_{scatt,1}$ is the relaxation time and $\kappa_1$ is the thermal conductivity in material 1. $\beta_{21}$ has a similar definition. To be consistent with the 2D conduction, these fractions are also calculated in 2D. By plugging the AMM expression into Eq. 25, the conductance $G_{neq}$ of this model has been shown to compare well with molecular dynamics simulations [39].

### E. Numerical implementation

The simulation allows us to compute the average flux density at a given position $x_0$ including the contributions of all waves:

$$\overline{P}_x(x_0) = \frac{1}{y_{\max} - y_{\min}} \int_{y_{\min}}^{y_{\max}} P_x(x_0, y) \, dy \quad (23)$$

with $y_{max}$-$y_{min}$ being a multiple of $a$, and $P_x(x_0,y)$ is defined in Eq. 5. The frequency-dependent transmission coefficient can be calculated as the ratio between the sum of transmitted heat flux intensities and the sum of the incident heat flux intensities $\overline{P}_0$:

$$\tau(\omega) = \frac{\int_{\theta_i} \left[ \int_{\theta_t} \overline{P}_t(\omega, \theta_i, \theta_t) \cos\theta_t \, d\theta_t \right] d\theta_i}{\int_{\theta_i} \overline{P}_0(\omega) \cos\theta_i \, d\theta_i} \quad (21)$$

where $\overline{P}_0(\omega, \theta_i)$ and $\overline{P}_t(\omega, \theta_i, \theta_t)$ are the average flux densities, respectively of the incident wave (identical for all incident angles $\theta_i$) and of the transmitted wave in a given angle $\theta_t$ for an incident angle $\theta_i$. Numerically, we extract $\overline{P}_{t,x}(\omega, \theta_i) = \int_{\theta_t} \overline{P}_{t,x}(\omega, \theta_i, \theta_t) \cos\theta_t d\theta_t$, where the index $x$ stands for the projection on the $x$ axis. The frequency-dependent coefficient is computed by discretizing Eq. 21:

$$\tau(\omega) \approx \frac{\sum_{\theta_i} \overline{P}_{t,x}(\omega, \theta_i) \Delta \theta_i}{\overline{P}_0(\omega)}. \quad (22)$$

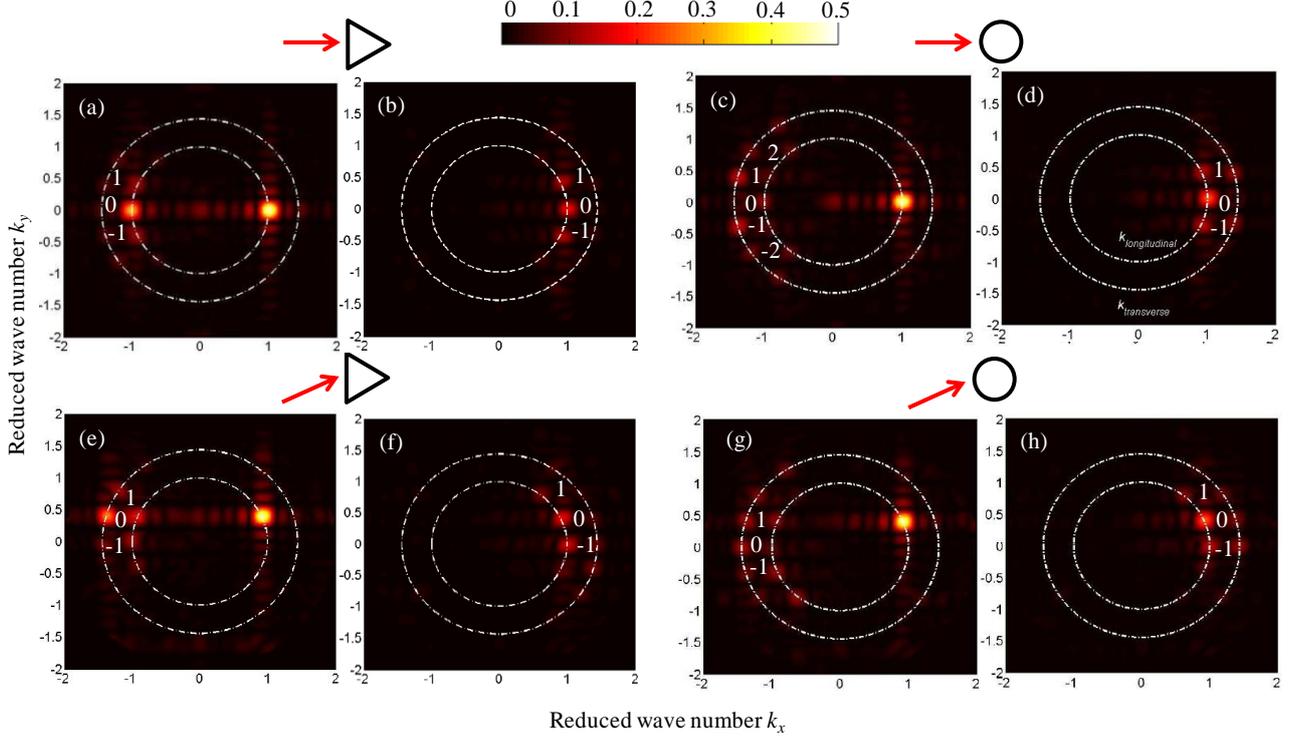

FIG 5. Maps of spatio-temporal TF in reciprocal space for $N_\lambda = 2.5$ in the case of circular holes (a-b-e-f) and in the case of equilateral triangular holes (c-d-g-h) : (a-c) incident region and (b-d) transmitted region for a wave in normal incidence ($\theta = 0$); (e-f) incident region and (g-h) transmitted region for a wave in oblique incidence with $\sin\theta = 0.4$. The dashed lines represent the iso-wavevector curves: the inner circle for the longitudinal one, the outer circle for the transverse one. The orders of diffraction are also shown in the maps (0 for the zeroth order, ±1 for the first order, ±2 for the second order).

## III. Results
### A. Diffraction effect – Spatial filtering

To analyze the transmission and reflection processes, we use the spatio-temporal Fourier Transform (FT) [41]. It has been used to investigate the dispersion relation of surface acoustic modes [42], acoustic band structure [43] and to observe the diffraction angles [44]. In infinite space and time domains, the FT is written as

$$U(\vec{k},t) = \int_{-\infty}^{\infty}\int_{-\infty}^{\infty} u(\vec{r},t) e^{i(\omega t - \vec{k}\vec{r})} d\vec{r}\, dt \,. \quad (27)$$

By limiting to a given spatial domain and integrating over one period $T = 2\pi/\omega$, the spatio-temporal FT becomes

$$U(\vec{k},t) = \frac{1}{l_x}\frac{1}{l_y}\frac{1}{T}\int_{t_1}^{t_1+T}\int_{x_0-l_x/2}^{x_0+l_x/2}\int_{y_0-l_y/2}^{y_0+l_y/2} u(x,y,t) e^{i(\omega t - k_x x - k_y y)} dx\, dy\, dt \,; \quad (28)$$

$[x_0 - l_x/2, x_0 + l_x/2]$ and $[y_0 - l_y/2, y_0 + l_y/2]$ are the bounds of the analyzed domain in $x$ and $y$ directions. We consider a time $t_1$ for which the stationary regime is already established, $l_y$ proportional to $a$, $x_0$ at left or right of the scattering elements with $l_x$ such that the addressed area is in the far field of the scattering elements. Both $u_x$ and $u_y$ components are analyzed with FT. In order to examine the conversion of polarization, we combine these two components to determine the relative contributions of the longitudinal and the transverse modes. The longitudinal and transverse amplitudes are obviously calculated with

$$\begin{cases} u_l = u_x \cos\theta + u_y \sin\theta \\ u_t = -u_x \sin\theta + u_y \cos\theta \end{cases} \quad (29)$$

where $\theta$ is the angle between $x$-axis and the wavevector $\vec{k}$.

Fig. 5 shows the maps of spatio-temporal FT of the displacement field in reciprocal space with $N_\lambda=2.5$, i.e. $\lambda=a/2.5$. The independent propagation of longitudinal $u_l$ and transverse $u_t$ modes are shown for two cases: (a-b-e-f) circular holes, (c-d-g-h) triangular holes. The two dotted circles represent the iso-wavevector curves for longitudinal waves ($k_l = \omega/v_l$) and for transverse waves ($k_t = \omega/v_t$). As the longitudinal velocity $v_l$ is larger than the transverse one $v_t$, the inner circle corresponds to $k_l$ and the outer one corresponds to $k_t$. In this figure, the wave vectors are normalized by the maximal longitudinal one.

In Fig. 5a and 5c, the incident waves in normal incidence ($\theta = 0$) are represented by the points on the longitudinal circle, with $k_x > 0$ and $k_y = 0$; this shows the propagation in the positive direction. On the same circle, the points of negative $k_x$ represent the diffracted waves: the centered point is associated to the reflected wave (order 0), the next two symmetric points, which have $k_y/k = 1/2.5 = 0.4$, are associated to the first order of diffraction, while the two outer points with $k_y/k = 2/2.5 = 0.8$ are associated to the second order of diffraction. These ratios are exactly the sine values of diffracted angles $\theta_1$ and $\theta_2$, as predicted by Bragg's law.

In addition, diffracted waves are also represented on the transverse circle. This shows the effect of conversion of polarization due to the arrays. In the transmitted region, diffracted waves that travel through the single array are represented with $k_x > 0$ in Fig. 5b and 5d. According to Snell's law, these transverse waves have the same $k_y$ component than the longitudinal wave of identical diffraction order. For an oblique wave with incidence such that $\sin\theta = 0.4$, the incident waves in Fig. 5e and 5g are found on the iso-longitudinal wavevector circle according to $k_y/k_l = 0.4$. As a consequence, the zeroth-order diffracted waves also have this ratio. In addition, higher orders are shifted of $n\lambda/a$ on the $k_y$-axis. The analysis in the two regions remains similar to the normal incidence case. While the amplitude of each diffracted wave depends on the hole shape, Bragg's law and Snell's law apply for all structures. Diffraction plays an important role on the propagation direction, leading to spatial filtering effect. Note that for long wavelengths ($\lambda > a \leftrightarrow N_\lambda < 1$) the propagation direction of the transmitted wave is the same as the one of the incident wave.

### B. Transmission coefficient and 2D thermal conductance

#### 1. Transmission coefficient $\tau(\omega)$ through the periodic array

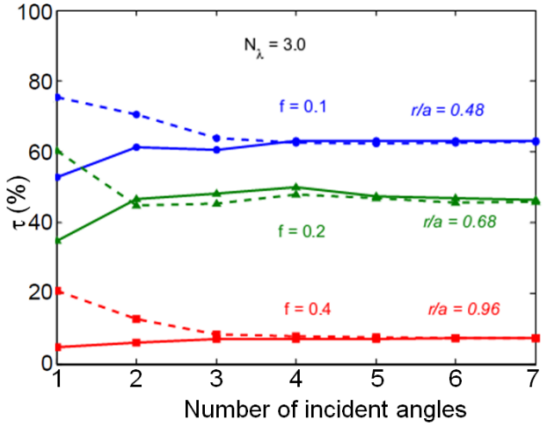

FIG 6. Frequency-dependent transmission coefficient $\tau(\omega)$ as a function of the number of incident angles for $N_\lambda = 3.0$ in two cases: (i) incidence toward the triangle bases (solid lines) and (ii) toward the vertices (dashed lines), both for a single array of equilateral triangles.

First, we analyze the frequency-dependent transmission coefficients as a function of the number of incident angles. As the coefficient depends on cosine values, the large angles, especially those near $\pi/2$, are less important than the small ones. In our simulations incident angles are characterized by sine values in [0–0.89]. Fig. 6 shows the frequency-dependent coefficients as a function of number of incident angles for the case $N_\lambda = 3.0$. In this example, the sine values included are 0, 0.11, 0.25, 0.4, 0.53, 0.67 and 0.85. The convergence of the coefficient value appears to be obtained for a rather small number of angles. We have verified that increasing this number up to 18 does not significantly improve the calculations in our case. In the following, we calculate the total transmission coefficients and the thermal conductance with 7 angles for each frequency.

(a)
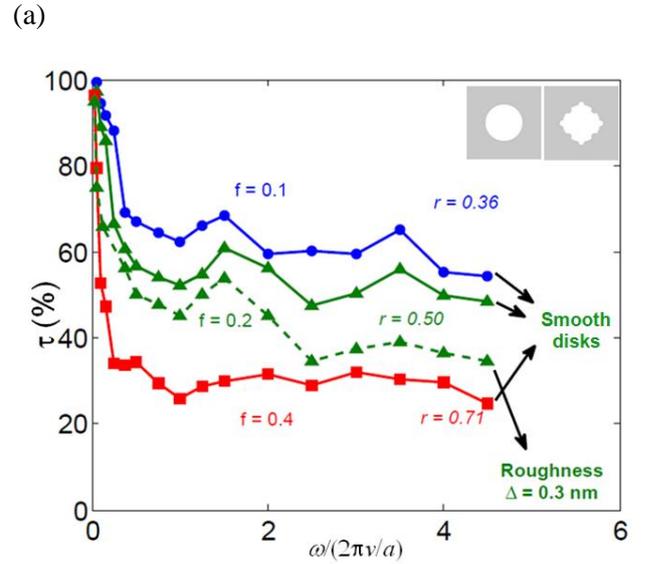

(b)
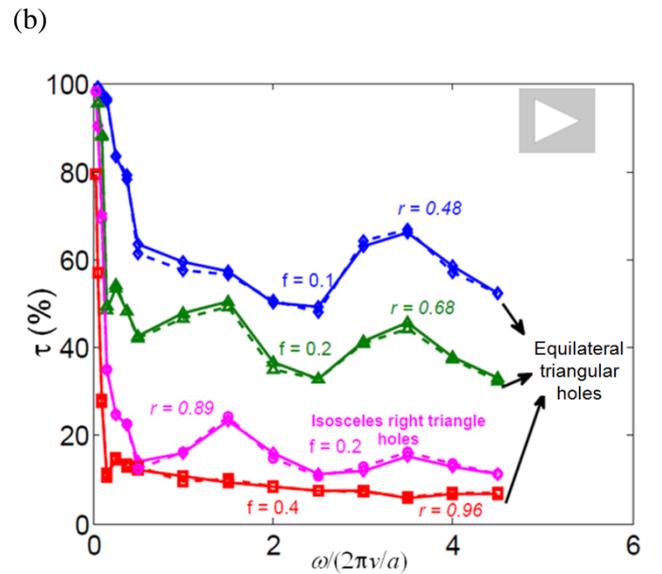

FIG 7. Frequency-dependent transmission coefficients $\tau(\omega)$ for arrays of (a) circular holes and (b) equilateral and right isosceles triangular holes (plain lines). In (a), the dashed line represents

the results for rough circular holes. The filling factor $f$ and the blocking ratio $r$ are shown for each curve.

We observe in Fig. 6 (and Fig. 7b) that the frequency-dependent transmission coefficients associated to heat fluxes impinging the bases or the vertices of the asymmetric triangles are equal when all propagation directions are included. This is a manifestation of reciprocity, which is fulfilled in our lossless and linear acoustic system (see more in [32]). Indeed no rectification can be observed in absence of non-linear mechanism. This is different to the case of acoustic waves only excited and observed in normal incidence [44, 45]. We also note in Fig. 6 that the frequency-dependent transmission coefficients depend strongly on the filling factor $f$: as an example, the lower the filling factor ratio, the larger the transmission.

Fig. 7 shows the frequency-dependent transmission coefficients as a function of the frequency $\omega/(2\pi v/a) = N_\lambda$ for (a) circular holes and (b) equilateral and right isosceles triangular holes. For all the presented longitudinal cases, we observe that the curves have same trend. The transmission coefficients reach unity at low frequency, and then reduce to a certain value when increasing the frequency. Two peaks at $N_\lambda = 1.5$ and $3.5$ are observed, while the non-zero minimum values take place for $N_\lambda = 2.5$. However, the values of the coefficient remain quite close for each filling factor/blocking ratio for frequencies larger than $N_\lambda = 0.5$ and we do not observe sharp features in the spectrum. This may be due to the "thermal" averaging due to the integration over all the excited angles. For the circular holes of Fig. 7a, the transmission coefficients are around 65% for $f = 0.1$ ($r = 0.36$), then around 55-60% for $f = 0.2$ ($r = 0.50$), and finally around 30% for $f = 0.1$ ($r = 0.48$). The same order is obtained for equilateral triangular holes, the transmission coefficients are the largest for $f = 0.1$ ($r = 0.48$), being close to only 60%. This means that phonons are already efficiently blocked and diffracted for a modest density of scatterers. The values of the coefficients are around 40% for $f = 0.2$ ($r = 0.68$), then less than 10% for $f = 0.4$ ($r = 0.96$). The same trend is observed for transverse waves (not shown here). Comparing the total coefficients with $f = 0.2$, they are different for each shape. The values of the frequency-dependent transmission coefficients decrease with the following order: disk, equilateral and right isosceles holes, in agreement with the values of the associated blocking ratios. Considering two cases with close values of $r$ but different values of $f$ - (i) circular array with $f = 0.2$, $r = 0.50$, and (ii) triangular array with $f = 0.1$, $r = 0.48$ - we observe close transmission coefficients through these arrays, around 60%. Despite the fact that wave scattering is often related to the area associated to the scatterer, the blocking ratio may be also a convenient way to describe the transmission.

Let us note that we have considered ideal shapes until now, neglecting possible roughness on the walls of the holes. In contrast to many other works, such roughness can be accounted for by designing directly complex shapes without relying on approximate analytical expressions. For example, a roughness of $\Delta=0.3$ nm was introduced to the case $f = 0.2$ studied in Fig. 7a by extruding half disks and grafting them between the extruded area (see inset in Fig. 7a). Note that such roughness is not random but identical on each hole if no supercell is considered. A strong decrease of the transmission can be observed, reaching 17%. Roughness especially impacts the transmission coefficients at high frequencies when short wavelengths become comparable to the roughness characteristic size. In the following, we restrict our study to smooth shapes.

### 2. *Phononic thermal conductance*

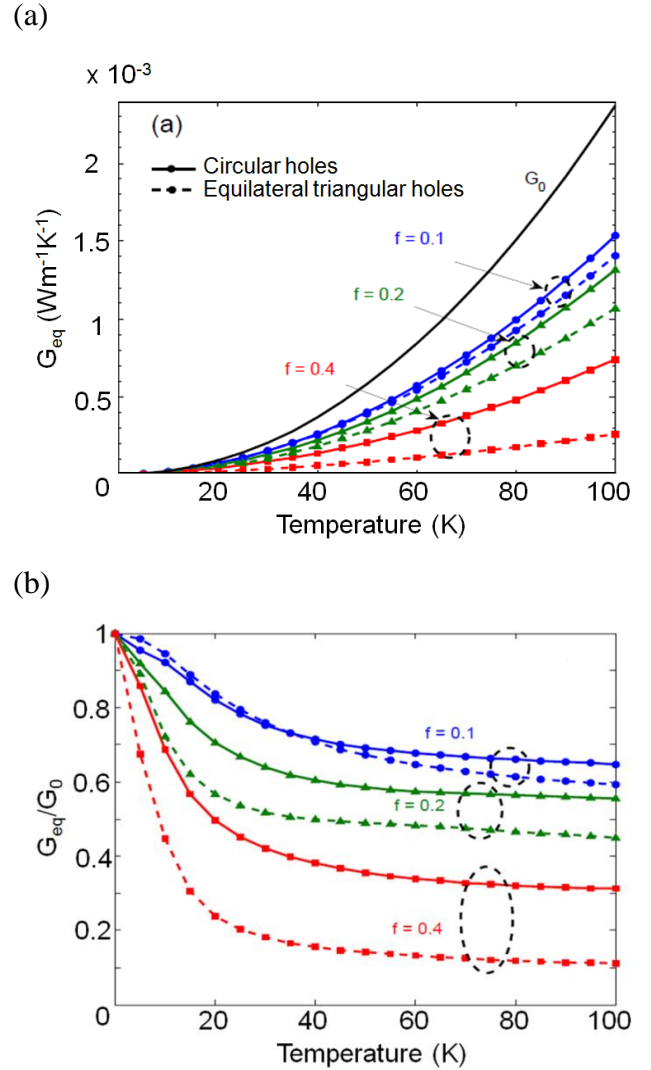

FIG 8. (a) Equilibrium phononic thermal conductance as a function of temperature in absence of hole array $G_0$ (solid black line), in presence of an array with circular holes (solid lines with symbols) and with equilateral triangular holes (dashed lines with symbols). (b) Ratio between the thermal conductance in presence of the array and without it. Blue lines for filling factor $f$ = 0.1, green lines for $f$ = 0.2, red lines for $f$ = 0.4.

The single array of periodic holes acts as a thermal barrier to which a thermal conductance can be associated. In order to calculate the thermal conductance as described in Section II.D, we consider temperatures exciting thermal wavelengths which are commensurate with the geometric parameters of the array such as its periodicity. Here, a periodicity $a$ = 10 nm is chosen as an example. The frequency dependence of the maximal monochromatic thermal conductance defined in Eq. 20 (section II. D) is considered. The thermal frequencies associated to temperatures ranging from 5 K to 100 K are well located in the simulated frequency band $\omega/(2\pi v/a)$ = [0.025 – 4.5]. We have checked that the summation of monochromatic thermal conductances $g_{\omega,max}$ with our discretization fills over 90% of the values of the integral for each temperature, as is customary for bosons which possess broad spectra. Note that the time-domain simulations are computer intensive (few hours of CPU time on a 12-core processor with RAM 64GB for one simulation) and that the final computation requires summation over frequency and angle.

The equilibrium thermal conductance $G_{eq}$ is shown in Fig. 8a for circular and triangular hole-based single arrays. The 2D conductance without the presence of hole $G_0$ (integral of Eq. 20) is also plotted for comparison The thermal conductance decreases when the filling factor increases, as the transmission coefficients are reduced (see Eq. 19). Moreover, for the same filling factor the thermal conductance of the triangular array is always smaller than that of the circular one. We remark that the temperature dependence of the thermal conductance in all cases is quadratic, i. e. $G_{eq} \propto T^2$: the modulation of the Bose-Einstein factor by the transmission coefficients does not lead to a different thermal behavior. The reduction of thermal conduction in the presence of single arrays normalized to $G_0$ is plotted in Fig. 8b. With $f$ = 0.1, the relative conductance reduces to 65%-62% for the two hole types, while the conduction drops to 30% and 10% with $f$ = 0.4.

To obtain the non-equilibrium conductance (Eq. 25), we take into account phonon-phonon Umklapp scattering. Our purpose is more to analyze how non-equilibrium affects the values of the phononic thermal conductance than to properly account for phonon volume scattering. Slack and Galginaitis [46] suggested the following form for the Umklapp process:

$$\tau_U^{-1}(\omega) = B_U \omega^2 T e^{-\theta_D/3T} \qquad (30)$$

where $B_U$ is a fitted parameter and $\theta_D$ is the Debye temperature. We consider the values of parameters calibrated to reproduce the experimental thermal conductivity of silicon in Ref. [47]. They are $\theta_L$ = 586 K and $B_U^L = 5.5 \times 10^{-20}$ $s^{-1}K^{-3}$.

The results are shown in Fig. 9 for the case of longitudinal waves propagating through an array of triangular holes. As the same material is present on the two sides of the array, one can define only one fraction $\beta$ = $\beta_{12}$ = $\beta_{21}$. Due to reduction of the transmission coefficient when widening the holes, the same trend is obtained for $\beta$ for temperatures from 5K to 100K. The non-equilibrium conductances are always higher than the equilibrium ones for all cases (see Eq. 25). The shift of the non-equilibrium conductance with respect to the equilibrium one increases with the transmission coefficient, i.e. when reducing the element size. The main conclusions of the preceding sections stay valid.

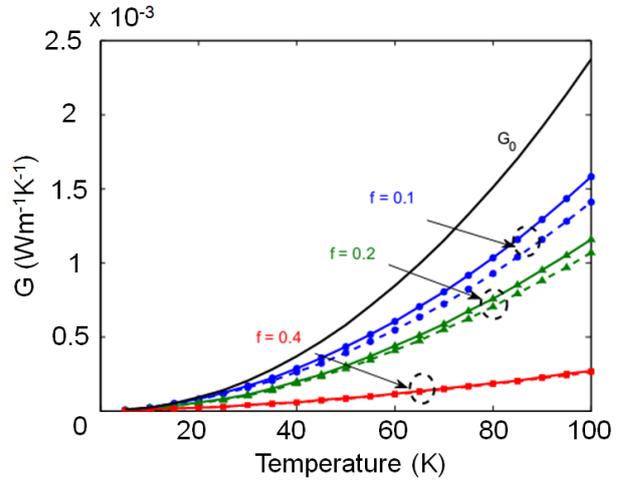

FIG 9. Phononic thermal conductance as a function of temperature: in absence of hole array $G_0$ (solid black line – equilibrium case for reference), in presence of a triangular array at equilibrium (dashed line with symbols) and out of equilibrium (solid lines with symbols). Blue lines for filling factor $f$ = 0.1, green lines for $f$ = 0.2, red lines for $f$ = 0.4.

## IV.  Conclusion

In summary, we have investigated the transmission of heat though a finite periodic array of scatterers, here holes, by solving the elastic wave equation in finite geometries. The acoustic waves, which model thermal phonons, have been excited in various directions to model the thermal "emission" by a heat source. By analyzing the spatio-temporal FT of the displacement fields, we observed (i) that the periodicity follows Bragg's law of diffraction, and (ii) that the conversion of polarization takes place according to Snell's law. The amplitude of the transmission of each phonon mode varies as a function of the scattering object shape, and we found that there is strong filtering effect even for modest size of the elements, such as for an equivalent filling ratio $f$ = 0.1. In particular, we observed that rough scatterers lead to lower transmission than smooth ones.

The frequency-dependent transmission coefficients of three types of arrays were found to have similar trends when varying the frequency. We noticed that the "blocking ratio" appears useful to determine if phonons can be transmitted: despite of the different shapes, the frequency-dependent transmission coefficients through objects of similar ratios are very close. Furthermore, for triangular (asymmetric) objects, the frequency-dependent coefficients associated to the heat flux impinging (i) the bases and (ii) the vertices have the same values. This is a consequence of reciprocity. We also observed that there is no particular frequency at which sharp spectral features such as a total gap could appear.

We highlighted that the single array acts as a thermal barrier, in a way very-closely related to thermal (Kapitza) boundary conductances. The phononic thermal conduction in the 2D case was characterized for these structures for temperatures ranging from 5K to 100K, in the example of isotropic silicon. Due to the impact on the frequency-dependent transmission coefficient, we observed that the hole size strongly influences the thermal conductance. A reduction of 90% could be reached for a blocking ratio $r = 0.71$. We also noticed that the presence of the array does not change the temperature dependence of the equilibrium thermal conductance: it merely filters the whole spectrum once a certain frequency is reached. We finally considered the impact of non-equilibrium close to the array.

This work is an important step related to the study of the wake-like phonon scattering properties through artificial interfaces, which are created by adding a periodic structure which is finite in at least one dimension. It may provide a basis for future investigations dealing also with non-linear mechanisms [48, 49] expected to exhibit thermal rectification effects.

## Acknowledgements

This work has been supported by ANR project RPDOC NanoHeat and INSA BQR MaNaTherm. We acknowledge useful discussions with S. Merabia. P.O.C. also thanks F. Alzina, E. Chavez, J. Gomis and C. Sotomayor.

## Appendix A: Wave equation in absorbing region

We recall the derivation of the equation used for the absorbing zones [35]. In 1D, the elastic wave equation is written as

$$\rho \frac{\partial^2 u}{\partial t^2} + \nabla T = 0 . \quad (A1)$$

This equation can be expressed in the following form:

$$\rho \frac{\partial^2 u}{\partial t^2} + \nabla(-C \nabla u) = 0 . \quad (A2)$$

In the absorbing zones, we introduce a new mass density and new elastic constants while keeping the same acoustic impedance $Z = \sqrt{\rho C}$: $\rho_{abs} = \frac{\rho}{d}$, $C_{abs} = C \cdot d$. Then, we look for a solution with a plane wave form $\vec{u} = e^{-\alpha|\vec{r}|} e^{i(\omega t - \vec{k}\vec{r})}$ in the absorbing regions. We can take

$$d(\omega) = \frac{1}{\sigma + i\omega} . \quad (A3)$$

By combining Eqs. A2 and A3, we obtain

$$\rho(\sigma^2 - \omega^2)u + i \cdot 2\rho\sigma\omega u - \nabla(-C \nabla u) = 0 . \quad (A4)$$

Finally, Eq. A.4 can be rewritten in the real space as a third-order partial differential equation, or also as follows:

$$\rho\left(1 - \frac{\sigma^2}{\omega^2}\right)\frac{\partial^2 u}{\partial t^2} - 2\rho\sigma \frac{\partial u}{\partial t} + \nabla T = 0 . \quad (A5)$$

This linear equation, which mixes frequency and time, may appear as unusual, but it leads exactly to the same displacement field that the one of the third-order equation when a plane wave is excited at ω. It can be useful if one prefers only to solve second-order partial differential equations in the computational domain to avoid potential additional discretization requirements.

## Appendix B: Transmission through a perfect Si-Ge interface

In this section, we compute the transmission through a perfect Si-Ge interface and compare the results with those obtained semi-analytically within the acoustic-mismatch model (AMM) [50]. This model captures the impedance mismatch effect of phonon transmission. At the interface, one part of wave is reflected and the other part is refracted at the other side of the interface following Snell's law:

$$\frac{\sin \theta_1}{v_1} = \frac{\sin \theta_2}{v_2} . \quad (B.1)$$

$\theta_1$ and $\theta_2$ are the incident and refraction angles, respectively. It is required that the incident angle be smaller than the critical angle $\theta_c = \operatorname{asin}(v_2/v_1)$. We consider longitudinal wave with velocity ratio $v_{Si}/v_{Ge} \approx 1.7$. As a consequence, the critical angle of waves propagating from Ge towards Si is $\theta_c \approx 35.°$, while there is no angle limit in the opposite direction.

Assuming that no inelastic scattering takes place at the interface, the transmission coefficient $t_{12}$ through a perfect interface between two media 1 and 2 is given in the AMM framework by

$$t_{12}(\omega,\theta_1) = \frac{4Z_1 Z_2 \cos\theta_1 \cos\theta_2}{(Z_1 \cos\theta_1 + Z_2 \cos\theta_2)^2} \text{ with } \theta_1 \leq \theta_c,$$ (B.2)

$$t_{12}(\omega,\theta_1) = 0 \text{ otherwise,}$$

where $Z_1$, $Z_2$ are the acoustic impedances of medium 1 and 2, respectively.

Fig. B.1 shows the transmission coefficients obtained for $N_\lambda = 1.25$ as a function of the angle of incidence for two cases: waves propagating from Si to Ge (Fig. B.1a), and from Ge to Si (Fig. B.1b). The coefficients calculated with the AMM are also plotted, in solid lines. The results are in good agreement with the AMM prediction in the [0-90°] range. This validates our method and allows us to study the transmission through the periodic array.

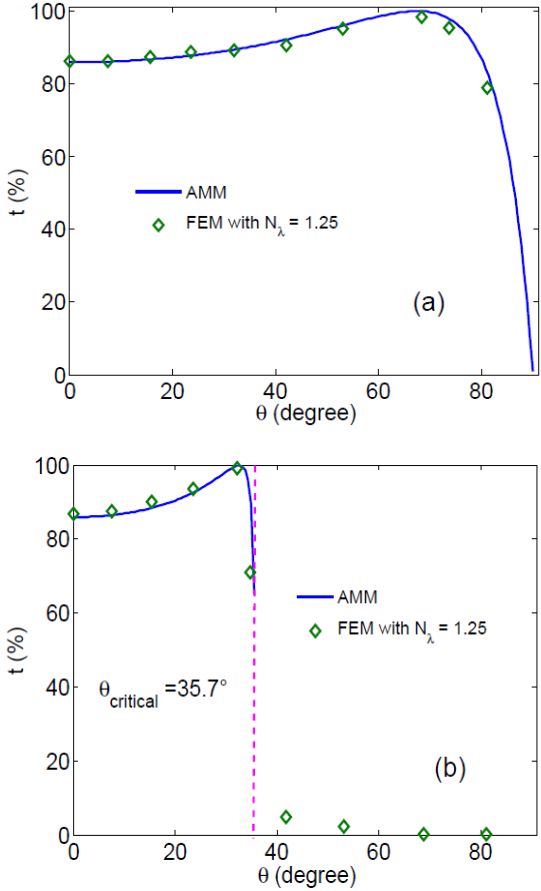

FIG B.1. Transmission coefficients as a function of the angle of incidence for waves crossing the Si/Ge interface (a) from Si to Ge, (b) from Ge to Si. The blue solid lines present the AMM calculation and symbols are for the results obtained from simulations with $N_\lambda = \frac{\omega}{2\pi v_{inc}/a} = 1.25$. The vertical dashed line shows the critical limit of 35.7° when the waves come from the Ge medium.

## Appendix C: Periodic conditions lead to diffraction

The periodic condition for wave propagation in the simulated domain, in particular for the oblique ones, is expressed in Eq. 11:

$$\sin\theta = \frac{n}{N_\lambda N_{rows}}.$$ (C.1)

### C.1. Propagation through a perfect interface

According to Snell's law, the refraction angle at the interface between two different materials is defined from the condition

$$\frac{\sin\theta_1}{v_1} = \frac{\sin\theta_2}{v_2},$$ (C.2)

that can be rewritten

$$\sin\theta_2 = \frac{v_2}{v_1} \sin\theta_1.$$ (C.3)

As a consequence,

$$\sin\theta_2 = \frac{v_2}{v_1} \cdot \frac{n}{N_{rows} N_{\lambda,1}} = \frac{v_2}{v_1} \cdot \frac{n}{N_{rows} \frac{\omega}{2\pi v_1/a}}$$
$$= \frac{n}{N_{rows} \frac{\omega}{2\pi v_2/a}} = \frac{n}{N_{rows} N_{\lambda,2}}.$$ (C.4)

This shows that the sine value of the refraction angle satisfies the periodic condition in medium 2. This is due to the well-known condition that the tangential component is conserved when crossing an interface.

### C.2. Propagation through a periodic array

The diffracted waves satisfy Bragg's law as expressed in Eq. 12:

$$a \sin\theta_m = m\lambda,$$ (C.5)

where $m$ is an integer which corresponds to the diffraction order characterized by the angle $\theta_m$ ($0 \leq m \leq [N_\lambda]$). By replacing (1) in (12), we obtain:

$$\sin\theta_m = m\frac{\lambda}{a} = m\frac{\frac{2\pi v}{\omega}}{a} = m\frac{2\pi v/a}{\left(\frac{2\pi v}{a}N_\lambda\right)}$$
$$= \frac{mN_{rows}}{N_\lambda N_{rows}}$$

(C. 6)

The numerator of Eq. C.6 is obviously an integer. By comparing Eq. 11 and Eq. C.6, we verify that the diffracted waves are satisfying the periodic conditions.